\documentclass[11pt]{article}
\usepackage{times}
\usepackage{sidecap}
\usepackage{geometry}
\usepackage{floatrow}
\usepackage{float}
\usepackage{hyperref}
\usepackage{graphics,graphicx}
\usepackage{subcaption}
\usepackage{epsf,wrapfig}

\usepackage{natbib,bm}

\usepackage{paralist}
\let\olditem\item
\renewenvironment{thebibliography}[1]{%
  \section*{\refname}
  \let\par\relax
  \renewcommand{\item}[1][]{\olditem[\textbullet]}%
  \inparaenum}{\endinparaenum}

\newcounter{aff}

\DeclareGraphicsExtensions{.jpg,.pdf,.png,.eps}

\usepackage{caption}
\usepackage{xcolor}

\usepackage{tcolorbox}

\bibpunct{(}{)}{;}{a}{}{,}

\usepackage{soul}

\definecolor{darkcyan}{rgb}{0.0, 0.0, 0.7}

\definecolor{boxcolor}{RGB}{193, 229, 255}
\definecolor{figboxcolor}{RGB}{193, 229, 255}
\definecolor{boxcolorgray}{RGB}{227, 227, 227}
\definecolor{darkcyan}{rgb}{0.0, 0.0, 0.7}

\sethlcolor{boxcolorgray}

\def\gtrsim{\lower 2pt \hbox{$\, \buildrel {\scriptstyle >}\over
{\scriptstyle \sim}\,$}}
\def\lesssim{\lower 2pt \hbox{$\, \buildrel {\scriptstyle <}\over
{\scriptstyle \sim}\,$}}

\def\approxlt{\lower.2em\hbox{$\buildrel < \over \sim$}}
\def\approxgt{\lower.2em\hbox{$\buildrel > \over \sim$}}

\def\arcsec{$^{\prime\prime}$}

\def\be{\begin{equation}}
\def\ee{\end{equation}}
\def\bea{\begin{eqnarray*}}
\def\eea{\end{eqnarray*}}

\def\arcsec{^{\prime\prime}}

\def\jref@jnl#1{{\rm#1}}
\def\aj{\jref@jnl{AJ}}			
\def\araa{\jref@jnl{ARA\&A}}		
\def\apj{\jref@jnl{ApJ}}			
\def\apjl{\jref@jnl{ApJ}}		
\def\apjs{\jref@jnl{ApJS}}		
\def\ao{\jref@jnl{Appl.~Opt.}}		
\def\apss{\jref@jnl{Ap\&SS}}		
\def\aap{\jref@jnl{A\&A}}		
\def\aapr{\jref@jnl{A\&A~Rev.}}		
\def\aaps{\jref@jnl{A\&AS}}		
\def\azh{\jref@jnl{AZh}}			
\def\baas{\jref@jnl{BAAS}}		
\def\jrasc{\jref@jnl{JRASC}}		
\def\memras{\jref@jnl{MmRAS}}		
\def\mnras{\jref@jnl{MNRAS}}		
\def\pra{\jref@jnl{Phys.~Rev.~A}}	
\def\prb{\jref@jnl{Phys.~Rev.~B}}	
\def\prc{\jref@jnl{Phys.~Rev.~C}}	
\def\prd{\jref@jnl{Phys.~Rev.~D}}	
\def\pre{\jref@jnl{Phys.~Rev.~E}}	
\def\prl{\jref@jnl{Phys.~Rev.~Lett.}}	
\def\pasp{\jref@jnl{PASP}}		
\def\pasj{\jref@jnl{PASJ}}		
\def\qjras{\jref@jnl{QJRAS}}		
\def\skytel{\jref@jnl{S\&T}}		
\def\solphys{\jref@jnl{Sol.~Phys.}}	
\def\sovast{\jref@jnl{Soviet~Ast.}}	
\def\ssr{\jref@jnl{Space~Sci.~Rev.}}	
\def\zap{\jref@jnl{ZAp}}			
\def\nat{\jref@jnl{Nature}}		
\def\iaucirc{\jref@jnl{IAU~Circ.}}
\def\aplett{\jref@jnl{Astrophys.~Lett.}}
\def\apspr{\jref@jnl{Astrophys.~Space~Phys.~Res.}}
\def\bain{\jref@jnl{Bull.~Astron.~Inst.~Netherlands}}
\def\fcp{\jref@jnl{Fund.~Cosmic~Phys.}}
\def\gca{\jref@jnl{Geochim.~Cosmochim.~Acta}}
\def\grl{\jref@jnl{Geophys.~Res.~Lett.}}
\def\jcp{\jref@jnl{J.~Chem.~Phys.}}	
\def\jgr{\jref@jnl{J.~Geophys.~Res.}}	
\def\jqsrt{\jref@jnl{J.~Quant.~Spec.~Radiat.~Transf.}}
\def\memsai{\jref@jnl{Mem.~Soc.~Astron.~Italiana}}
\def\nphysa{\jref@jnl{Nucl.~Phys.~A}}
\def\physrep{\jref@jnl{Phys.~Rep.}}
\def\physscr{\jref@jnl{Phys.~Scr}}
\def\planss{\jref@jnl{Planet.~Space~Sci.}}
\def\procspie{\jref@jnl{Proc.~SPIE}}

\def\arcsec{\hbox{$^{\prime\prime}$}}

\def\ion#1#2{#1$\;${\small\rm\@Roman{#2}}\relax}

\newcount\lecurrentfam
\def\LaTeX{\lecurrentfam=\the\fam \leavevmode L\raise.42ex
\hbox{$\fam\lecurrentfam\scriptstyle\kern-.3em A$}\kern-.15em\TeX}
\def\sizrpt{
(\fontname\the\font): em=\the\fontdimen6\font, ex=\the\fontdimen5\font
\typeout{
(\fontname\the\font): em=\the\fontdimen6\font, ex=\the\fontdimen5\font
}}
\definecolor{dark}{rgb}{ 0.34, 0.09, 0.49}

\def \Cii{[C\textsc{ii}]}
\def \Ci{[C\textsc{i}]}

\usepackage{fancyhdr,url}
\usepackage{transparent}
\usepackage{tablefootnote}
\usepackage{mathpazo}

\begin{document}

\noindent
\begin{raggedright}
{\Huge\bfseries ALMA2040: Understanding the Early Assembly of Galaxies \\[0.5em]
\Large The Formation of Star-Forming Clumps and their Role in Shaping Galaxy Evolution
}\end{raggedright}\\

\noindent

Anita Zanella$^{\dagger,1}$,
Andreas L. Faisst$^{\dagger,2}$, 
Rodrigo Herrera-Camus$^{\dagger,3,4}$,
Emily Chien$^{\dagger,5}$,
Bel\'en Alcalde Pampliega$^{6}$,
Manuel Aravena$^{3,7}$,
Cecilia Bacchini$^{8}$,
Tom Bakx$^{9}$,
Andrea Bolamperti$^{10,11}$,
Leindert A. Boogaard$^{12}$
Marusa Bradac$^{13}$,
Carla Cornil-Ba\"{i}otto$^{3,14}$,
Pierre Cox$^{15}$,
Miroslava Dessauges-Zavadsky$^{16}$,
Tanio Diaz-Santos$^{17,18}$,
Jeremy Fensch$^{19}$,
Jonathan Freundlich$^{20}$,
Mahsa Kohandel$^{1}$,
Andrea Lapi$^{21}$,
Umberto Maio$^{22,23}$,
Juan Molina$^{3,14}$,
Dieu D. Nguyen$^{24}$,
Miguel Pereira-Santaella$^{25}$,
Francesca Rizzo$^{26}$,
Justin Spilker$^{27}$,
Hannah Stacey$^{28}$,
Livia Vallini$^{1}$,
Enrico Veraldi$^{21}$

\begin{tcolorbox}[boxsep=3pt,left=0pt,right=0pt,top=1pt,bottom=1pt,colframe=boxcolor,colback=boxcolor]
\justify
\textbf{Abstract:} 
Astrophysics seeks to explain how galaxies form and evolve through the interplay of stars, gas, and dust. Giant molecular clouds (GMCs), the main reservoirs and sites of star formation, link the cold interstellar medium to the birth of stars and stellar structures. Their properties reflect galactic disk conditions, and their evolution is shaped by stellar feedback and host-galaxy dynamics. Studying GMCs across cosmic time is therefore key to understanding how gas turns into stars and drives galaxy evolution.
{\em Hubble Space Telescope} and {\em James Webb Space Telescope} observations reveal compact star-forming clumps in galaxies from Cosmic Noon to the Epoch of Reionization ($z\approx6$), but their link to the molecular gas reservoir is still unclear. We do not yet know how GMC properties vary with redshift and environment, how molecular gas, dust, and young stars are arranged, or whether clumps form in unstable disks or disturbed systems. Addressing these questions requires spatially resolved observations of the cold interstellar medium at the scales probed by optical/near-infrared facilities.
{\em ALMA2040}, a transformational next-generation sub-mm facility, will provide the sensitivity and angular resolution needed to detect individual GMCs in representative lensed and unlensed galaxies at $z>1$. CO and \Cii, complemented by dust continuum and molecular-gas kinematics, will measure the molecular gas reservoir, its physical conditions, and its dynamical state. Together with {\em JWST} and the ELT, {\em ALMA2040} will connect cold gas, dust, and stellar structures across cosmic time.
{\em ALMA2040} will revolutionize our understanding of
{\em (i)} the properties and evolution of GMCs and their connection to star-forming clumps,
{\em(ii)} the spatial interplay between molecular gas, dust, and young stars, and
{\em (iii)} the dynamics, survival, and formation of GMCs and stellar clumps.

\end{tcolorbox}

\newpage{}

\section{Scientific context and open questions}
\label{sec:context}

Understanding how galaxies form and evolve requires connecting their gas reservoirs to the formation of stars and the emergence of internal structure across cosmic time. Giant molecular clouds (GMCs) are central to this picture: they form from cooling and condensation of accreted gas, provide the sites of star formation, and are regulated by the gravitational state of galactic disks and by stellar feedback \citep{Toomre1964, Chevance2020,Dekel2009}. In the local Universe, GMCs typically have sizes of $5-100,\mathrm{pc}$ and molecular gas masses of $\mathrm{M_{H_2} \sim 10^4 - 10^7\,M_\odot}$ \citep{Bolatto2008, Miville-Deschenes2017, Freeman2017}.

HST and JWST have revealed abundant compact, star-forming structures from Cosmic Noon ($z\approx2$) to the Epoch of Reionization \citep[$z\approx6$; e.g.,][]{Conselice2014, Huertas-Company2023, Faisst2026b, HerreraCamus2025} and even beyond \citep{Fujimoto2025, Nakane2025, Abdurrouf2025, Messa2026}. Typical stellar clumps have stellar masses of $\mathrm{M_\star \approx 10^7 - 10^9\,M_\odot}$ and sizes below $\sim1\,\mathrm{kpc}$ \citep[e.g.,][]{Guo2015, Zanella2015, Kalita2023}. Strong gravitational lensing has extended this view to smaller and fainter structures: increasing magnification reveals progressively smaller clumps, with strongly lensed systems reaching $\mathrm{R_e \approx 10 - 100\,pc}$ and $\mathrm{M_\star \approx 10^6 - 10^8\,M_\odot}$, approaching individual stellar-cluster scales comparable to the structures observed at lower redshifts \citep{Livermore2015, Vanzella2017a, Cava2018, Vanzella2021, Vanzella2022, Mestric2022, Messa2022, Claeyssens2023, Adamo2024, Messa2024, Claeyssens2025, Claeyssens2026}.
Understanding whether the molecular clouds that fuel star formation are the direct precursors of the observed massive stellar clumps observed at high redshifts is therefore a key step toward linking the cold interstellar medium to star cluster and galaxy assembly.

Observations and simulations increasingly support a picture in which most clumps at $z\sim1-4$ form \emph{in situ} through gravitational fragmentation of gas-rich disks rather than being primarily merger products \citep{Shibuya2016, Zanella2019, DeLaVega2026,Escala2008}. If stellar clumps are the outcome of molecular cloud collapse ({\em i.e.,} GMCs), their formation should be reflected in the properties, distribution, and dynamics of the molecular gas. Yet direct observations of individual GMCs at $z>1$ remain rare. Existing resolved CO and \Cii~studies at $z\approx1-3$ have found clouds with masses, surface densities, and pressures substantially above those typical of nearby GMCs \citep{Dessauges-Zavadsky2019, Dessauges-Zavadsky2023, Zanella2024}, but current samples are small and dominated by the brightest clouds in strongly lensed galaxies.

ALMA2040 will overcome this selection limit by combining high angular resolution with $>5\times$ better line and continuum sensitivity than current ALMA. These observations will link GMCs, stellar clumps, and their host galaxies in representative samples, including modestly magnified and ultimately unlensed systems, to address three connected questions:
\begin{itemize}
    \item {\bf What sets the physical state and star-formation efficiency of GMCs at high redshift?} Measuring cloud masses, sizes, surface densities, excitation, and dense-gas fractions will establish how the molecular reservoir varies with redshift and environment, and whether the extreme conditions inferred today are typical in high-redshift galaxies.
    \item {\bf How do molecular clouds form and survive within the turbulent high-redshift ISM?} Resolved line kinematics will connect the dynamics of individual GMCs to the stability and motions of their host disks, testing whether clouds are bound, transient, or pressure-confined and how their lifetimes depend on galactic environment.
    \item {\bf How do GMCs produce the stellar clumps seen by JWST at high redshifts?} Joint measurements of gas, dust, and stellar structure will test whether stellar clumps are the products of molecular-cloud fragmentation and collapse, and how this connection changes from Cosmic Noon to the Epoch of Reionization.
\end{itemize}

In this white paper, we show that ALMA2040 is uniquely suited to address these questions. Its sub-mm capabilities will provide the molecular-gas and dust measurements that complement JWST and ELT observations of the stellar component, opening a direct view of the physical processes governing star formation and galaxy assembly across cosmic history.

\subsection{ALMA2040: science case}
\label{subsec:science_case}
\begin{wrapfigure}{r}{0.55\textwidth}\vspace{-17mm}
      \begin{center}
        \includegraphics[width=1\textwidth]{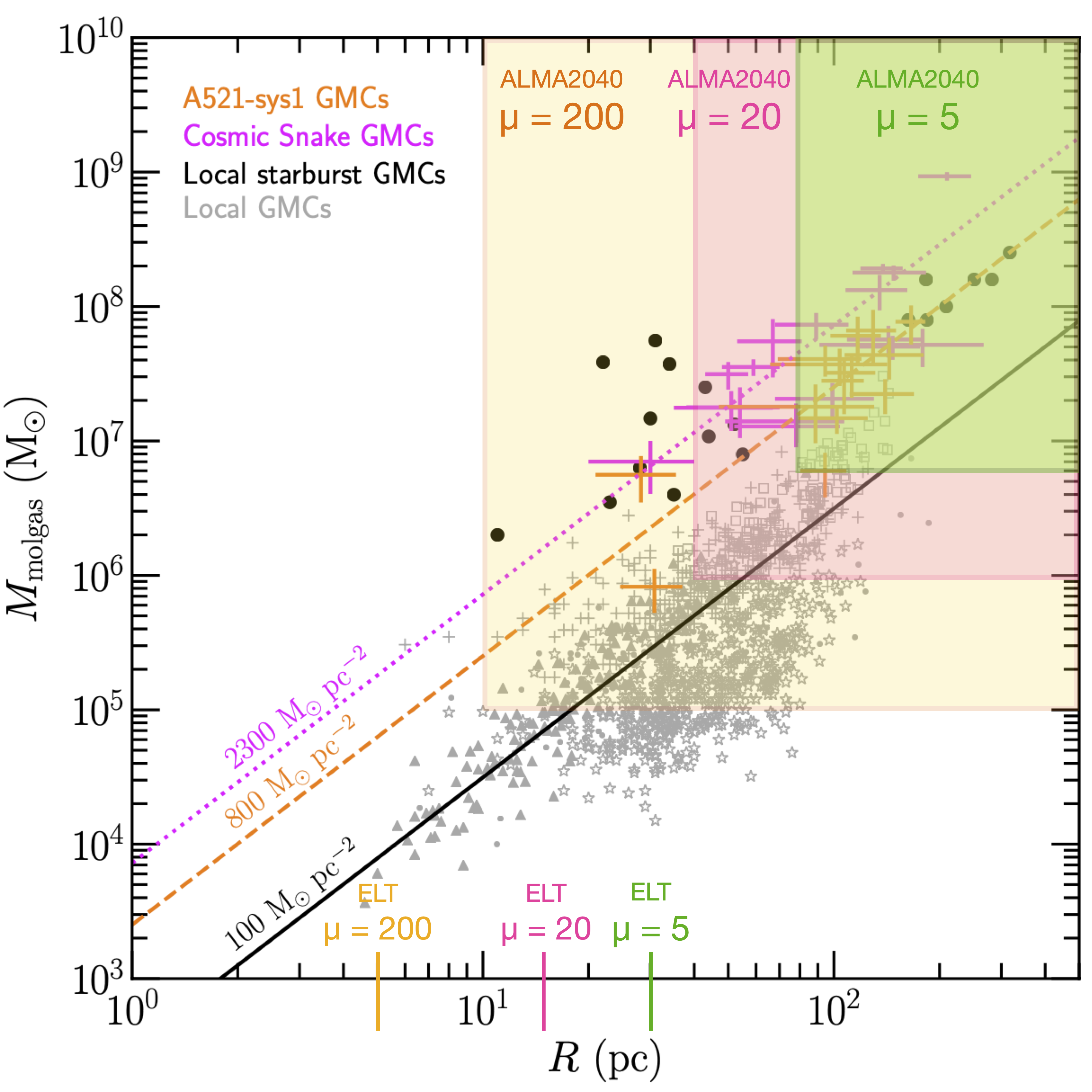}
    \end{center}
    \vspace{-\topsep} 
    \caption{Mass and radius of GMCs at $z \sim 1$ \citep{Dessauges-Zavadsky2019}, in local galaxies \citep{Bolatto2008, Heyer2009, Donovan_Meyer2013, Colombo2014, Corbelli2017}, and in nearby mergers/starbursts \citep{Wei2012, Leroy2015}. Colored regions indicate the parameter space spanned by ALMA2040 (depending on $\mu$ and considering $2 < z < 8$). The vertical lines on the $x$ axis shows the scales probed by ELT. The synergy of these two observatories will enable observations of high-redshift GMCs down to molecular gas masses and sizese comparable to local GMCs. Adapted from \cite{Dessauges-Zavadsky2023}. \vspace{-12mm}
    }
    \label{fig:GMCs}
\end{wrapfigure}
ALMA2040 will enable three complementary measurements in large, representative samples of star-forming, clumpy galaxies at $z > 1$:
(1) resolve individual GMCs comparable to the ones in local galaxies and measure their molecular gas masses and physical conditions;
(2) spatially resolve and map the dust continuum from galactic to GMC scales; and
(3) characterize molecular gas kinematics to connect cloud dynamics to the stability and evolution of their host galaxies.

\subsubsection{Characterizing GMCs across redshift and galactic environments}
\label{subsubsec:GMC_detection}

Molecular hydrogen ($\mathrm{H_2}$), the fuel for star formation, cannot be directly observed at high redshift, so CO is commonly used as its primary tracer \citep{Bolatto2013}.
While some resolved observations of CO emission of individual GMCs have been carried out \citep[e.g.,][]{Bolatto2008}, despite its fundamental importance, such measurements remain extremely scarce at $z>1$.
The first robust detections of individual GMCs in distant star-forming galaxies were obtained in the strongly lensed main-sequence systems Cosmic Snake and A521 at $z\approx1$. With $\lesssim0.2\arcsec$ ALMA observations, these studies resolved tens of clouds in CO with characteristic sizes of $\sim30-200\,\mathrm{pc}$ and molecular gas surface densities of $\Sigma_{\rm gas}\approx 10^3-10^4\,\mathrm{M_\odot\,pc^{-2}}$, about an order of magnitude above local values \citep[Fig~\ref{fig:GMCs};][]{Dessauges-Zavadsky2019, Dessauges-Zavadsky2023}. Such extreme properties are more similar to those of GMCs in local starbursts.
GMCs are often offset from rest-frame UV clumps, implying rapid disruption or dispersal after star formation starts on timescales of a few tens of Myr \citep{Dessauges-Zavadsky2023,ForsterSchreiber2011}, similar to GMC lifetimes in nearby galaxies \citep{Kruijssen2019, Chevance2020, Kim2021}. Whether such short lifetimes and extreme conditions are universal is unclear and requires resolving star-formation cycle phases at sub-kpc scales using multiple continuum and line tracers across diverse high-$z$ ISM conditions.

At higher redshift, only higher-$J$ CO transitions can be observed or CO observations as a whole become prohibitively expensive. This makes the far-infrared fine-structure line \Cii~$\lambda158\,\mu$m, one of the brightest far-infrared lines, an attractive alternative tracer of the cold molecular gas reservoir, although caution is advised as \Cii~is of multi-phase origin \citep{Madden1993,Heiles1994,Kim2002,Pineda2013,Tarantino2021}.
\Cii~is a dominant coolant of the cold ISM \citep{Stacey1991, Gracia-Carpio2011, Carilli2013, Decarli2025} and correlates with molecular gas content both observationally and theoretically \citep{Zanella2018, Madden2020, Dessauges-Zavadsky2020, Gururajan2023, Pallottini2017b, Ferrara2019, Sommovigo2021, Vizgan2022, Vallini2025}, although with a large scatter \citep{Benedetta2025,benedetta2024}.
Resolved \Cii~observations of a strongly lensed galaxy at $z\approx3.4$ already indicate a broad range of molecular gas depletion times, from $\approx0.16\,\mathrm{Gyr}$ to $\gtrsim0.65\,\mathrm{Gyr}$, suggesting substantial diversity in cloud-scale star-formation efficiency \citep{Hodge2015, Bethermin2023, Freundlich2013, Nagy2023}. However, these results come from only a handful of exceptionally bright and highly magnified systems, leaving it unclear how much of the apparent diversity is intrinsic.

ALMA2040 will turn these first detections into a population study and make it possible to:

\begin{enumerate}
\item Detect GMCs in galaxies with modest ($\mu\lesssim5$) or no gravitational magnification. This will mitigate the bias toward rare, highly magnified systems. Because the source-plane area with magnification greater than $\mu$ scales approximately as $\mu^{-2}$, moving from $\mu\gtrsim20$ to $\mu\lesssim5$ greatly enlarges the accessible population. Large samples will enable the statistical measurement of the intrinsic cloud-to-cloud and clump-to-clump scatter in various properties (mass, surface density, virial parameter, depletion time) and test how GMC and clump properties depend on stellar mass, star formation rate, environment, and cosmic time.

\item Detect more than an order-of-magnitude lower-mass clouds, comparable to local GMCs, especially in highly magnified systems, enabling the characterization of the GMC mass function Cosmic Noon and beyond and its direct comparison to nearby galaxies (Fig.~\ref{fig:GMCs}).

\item Extend resolved GMC studies to $z>3$ using both CO and \Cii, to test whether cloud properties, star-formation efficiencies, and depletion times evolve with redshift and to establish how the molecular reservoir relates to the stellar clumps characterized by JWST and ELT.

\item Measure the physical conditions of molecular gas through multi-$J$ CO and \Ci~observations. Modeling the CO spectral line energy distribution (SLED) has been essential for disentangling multi-component molecular ISM in local and Cosmic Noon galaxies \citep{Carilli2013, Daddi2015}. ALMA2040 will extend this approach from the most luminous systems to representative galaxy samples and, crucially, to individual clumps and GMCs. This will test whether the excitation and density distributions inferred locally persist at higher redshift, where GMC surface densities and pressures appear more extreme.

\item Measure the dense molecular component with tracers such as HCN, HCO$^+$, and CS. Their high critical densities directly probe the gas most closely associated with star formation, and their luminosities correlate with infrared luminosity and star formation rate in local galaxies \citep{Gao2004}. The first spatially resolved dense-gas detections in a typical main-sequence galaxy at Cosmic Noon \citep{Chen2026} show that such diagnostics are becoming feasible beyond starburst and AGN-dominated systems. ALMA2040 will extend them to clumps, to test whether the dense-gas fraction regulates clump-scale star-formation efficiencies and lifetimes, and how it varies with redshift and environment.
\end{enumerate}

Together, these observations will show whether the unusual GMC properties in early high-redshift detections are typical, pinpoint the parameters governing cloud-scale star formation, and track how the molecular reservoir evolves from the Epoch of Reionization to today.

\subsubsection{Resolving the dust distribution from galactic to GMC scales}
\label{subsubsec:dust_distribution}

Dust is an important component of galaxies and tightly connected to galaxy evolution. It not only affects the observed stellar emission by absorbing UV and optical radiation and re-emitting it in the far-infrared, it also plays a crucial role in cooling and therefore the formation of stars. Its formation and destruction furthermore depends on the local ISM condition and is tied to the star formation itself. It is therefore essential to understand its spatial distribution and relation to young stars, stellar population ages, UV and optical emission, hence molecular clouds. While global dust properties of high-redshift galaxies start to be probed, resolved dust measurements down to clump scales are still largely missing, leaving open key questions such as what are the dust content and properties of high-$z$ GMCs and what are the dust production mechanisms.

JWST now provides resolved rest-frame UV to mid-infrared attenuation, dust grain properties and stellar-population diagnostics at high redshift: spectroscopy constrains attenuation curves \citep{Markov2025, Shivaei2025}, the UV bump constrains dust composition \citep{Witstok2023,Nanni2025}, NIRCam imaging enables resolved maps of $A_V$ and $\beta_{\rm UV}$ \citep{Abdurrouf2023, Gimenez-Arteaga2024, Tanaka2024, Markov2026}, NIRSpec IFU data provide Balmer-decrement maps from H$\alpha$/H$\beta$ \citep{Parlanti2025, Mawatari2026,Tsujita2026,Faisst2026}. 
Recently, spatially resolved properties of dust molecules measured through polycyclic aromatic hydrocarbons (PAH) emission was obtained from MIRI IFU observations \citep{Wang2026}.
These observations reveal strongly structured dust attenuation, but they infer dust indirectly from the attenuated stellar light or on poor spatial resolutions.

Direct ALMA observations of the dust continuum instead remain sensitivity limited, with resolved measurements generally restricted to the brightest regions. Although dust temperatures and infrared luminosities are now measured for increasing numbers of high-redshift galaxies \citep{Inoue2020, Faisst2020a, Inami2022, Sommovigo2022, Algera2024, Bowler2024, algera2025}, dust maps at individual clump and GMC scales are largely inaccessible (but see \citealt{Cornil-Baioto2026} for a recent example). This matters because theoretical models predict that the relative geometry of dust and stars can strongly shape attenuation and infrared emission \citep{Behrens2018, Ma2019}. In particular, clumps are expected to have higher dust temperatures, larger $A_V$, stronger UV attenuation, and redder UV slopes than galaxy-integrated values \citep{Nakazato2026}, implying that unresolved attenuation corrections may bias the inferred properties of individual clumps.
Furthermore, GMCs are not only the cradles of star formation, but also dust production. Characterizing the lifetimes of such clouds, their dust masses, metallicity, and stellar ages therefore provides constraints on the early dust production mechanisms, which is directly relevant to the emergence of dust in very high redshift galaxies.

ALMA2040 will resolve the dust continuum on the same physical scales targeted by JWST, from GMCs ($\approx 10-100\,{\rm pc}$ in strongly lensed systems) to individual star-forming clumps in unlensed galaxies at Cosmic Noon and beyond. Critically, deriving dust temperature, dust mass, and optical depth at these scales cannot be achieved from a single continuum detection alone: Because dust temperature, emissivity index $\beta$, and mass are degenerate in single-band flux measurements, robust constraints require continuum detections across a wide frequency range that sample both the Rayleigh-Jeans and near-peak regimes of the modified blackbody SED. ALMA2040's simultaneous multi-band observing capability directly meets this requirement, delivering resolved continuum measurements across the full frequency range in a single observation rather than requiring separate, sequential band-by-band campaigns. Combining these multi-band resolved dust continuum measurements with resolved gas tracers (CO and [C II]) and stellar population SED fitting and spectroscopy from JWST and future ELT will allow a direct comparison of the distributions of dust, molecular gas, and young stars. This will constrain dust temperatures, masses, optical depths, and dust-to-gas ratios on GMC scales beyond the local Universe, and test how dust geometry and properties vary with redshift, environment, and local star formation activity.


\subsubsection{Constraining molecular gas motions and GMC survival}
\label{subsubsec:gas_motion}

The origin and evolution of stellar clumps remain uncertain, particularly at $z>4$, where the relative importance of mergers, rapid gas accretion, and disk instabilities may differ from Cosmic Noon \citep{Dessauges-Zavadsky2020,Romano2021,Faisst2025}. At $z\approx1-4$, observations and simulations broadly support {\em in situ} formation through fragmentation of gas-rich, turbulent disks, with a smaller contribution from interactions and mergers \citep{Bournaud2014, Shibuya2016, Mandelker2017, Zanella2019, DeLaVega2026}. Yet the molecular gas reservoir has rarely been mapped at comparable spatial resolution: most existing kinematic studies trace ionized gas or far-infrared lines \citep{Lelli2023, 2024Rizzo, Lee2025}. The key missing measurement is therefore the dynamical connection between the cold molecular phase, disk (in-)stability, and stellar clump formation. Building such a link would enble answering to crucial questions such as what is the origin of clumps (disk instability vs mergers) and what is the lifetime of high-redshift GMCs.

Spatially resolved molecular lines from ALMA2040 will provide rotation curves, velocity dispersions, velocity gradients, and molecular surface densities, distinguishing rotationally supported, dispersion-dominated (i.e., mergers), and dynamically disturbed systems. Combined with JWST or ELT stellar morphologies, these measurements will test whether clumps preferentially form in gravitationally unstable disks, dynamically disturbed systems, or both, and how this balance changes with redshift and galaxy properties.

The same data will determine whether individual GMCs are gravitationally bound, transient, or pressure-confined. Resolved GMCs in the strongly lensed Cosmic Snake and A521-sys1 at $z\simeq1$ already show masses, surface densities, and velocity dispersions substantially above local values; in A521-sys1, most clouds are consistent with approximate virial equilibrium \citep{Dessauges-Zavadsky2019,Dessauges-Zavadsky2023,Nagy2023}. These results suggest that GMC properties respond to the ambient ISM, but the limited sample prevents a test of this picture across cosmic time. To shed light on this open issue, a central diagnostic is the virial parameter ($\alpha_{\rm vir}$).
\begin{figure*}[t!]
\centering
\includegraphics[angle=0,width=0.75\textwidth]{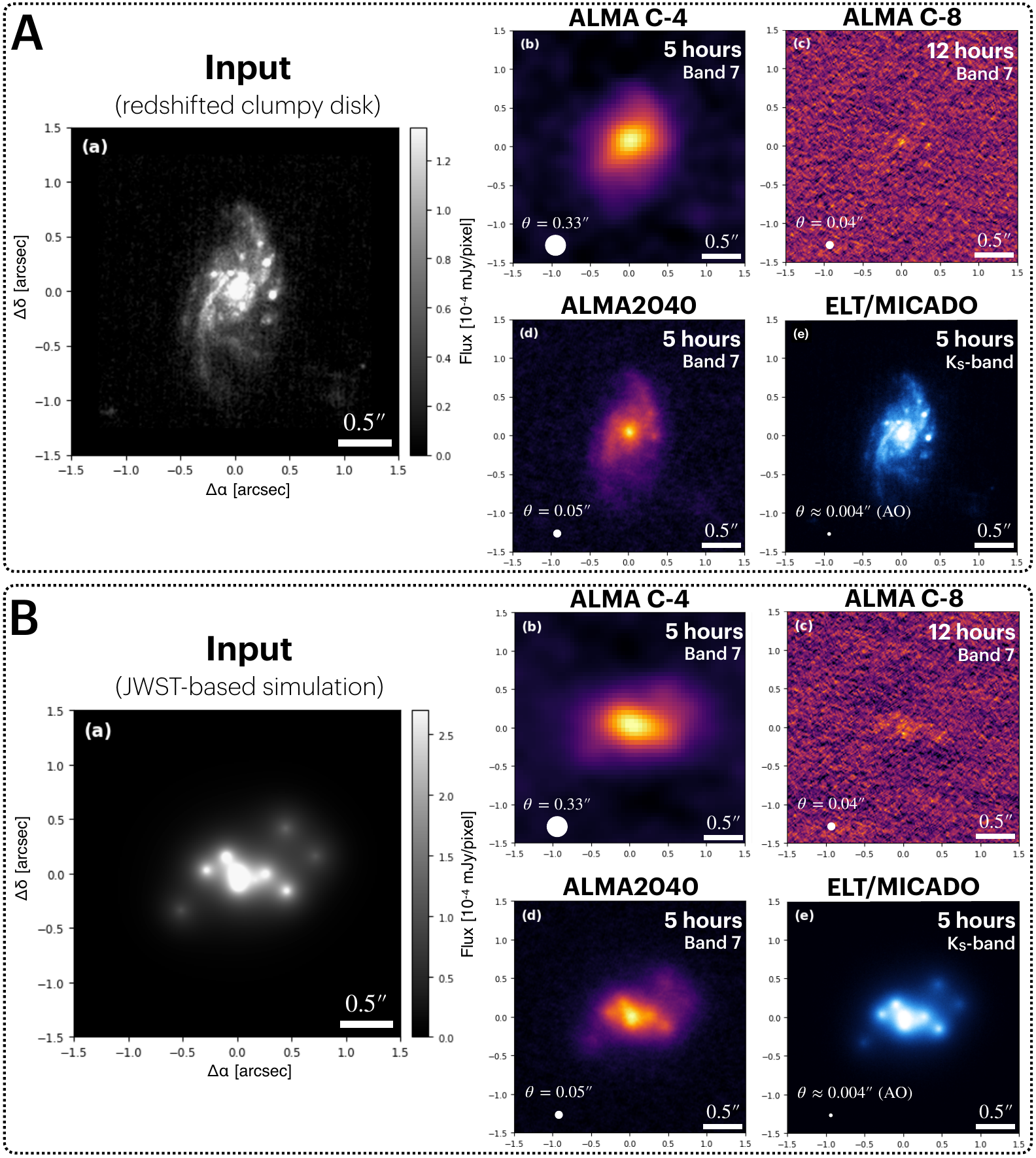}\vspace{-2mm}
\caption{
GHOST simulation of two galaxy models (A: based on an empirical clumpy galaxy at $z\approx1$, redshifted to $z=5$; B: based on observed clumpy galaxies at $z\approx5$ from large ALMA programs) in the present ALMA C-4 ({\em (b)}; $\theta\approx0.33\arcsec$) and C-8 ({\em (c)}; $\theta\approx0.04\arcsec$) configurations as well as fiducial ALMA2040 observations in the intermediate configuration ({\em (d)}; $\theta\approx0.05\arcsec$). The panels {\em (e)} show the corresponding simulation in $K_s$-band from ELT/MICADO with AO assistance (see text for more details).
\label{fig:ghost}
}
\end{figure*}

Nearby galaxies show a broad distribution of GMC virial parameters, typically $\alpha_{\rm vir}\approx1$ to a few, with substantial environmental variations \citep{Sun2018,Chevance2020}. ALMA2040 will extend such measurements to large samples over a broad range of redshift, stellar mass, star-formation rate, and environment. Cloud sizes, linewidths, and molecular masses will test whether the local and $z\approx1$ virial scaling relations persist to $z>4$, while providing an independent constraint on the CO-to-$\mathrm{H_2}$ conversion factor by comparing virial and CO-based masses for demonstrably self-gravitating clouds \citep{Bolatto2013,Dessauges-Zavadsky2023}.

GMC survival also depends on the host galaxy. Shear, gravitational collapse of the surrounding ISM, orbital and epicyclic perturbations, spiral-arm interactions, and cloud--cloud collisions can compress or disperse clouds on timescales set by the local dynamical environment \citep{Linden2017,Linden2021,Chevance2020,EspejoSalcedo2025}. Combining molecular-gas velocity fields with disk surface densities and rotation curves will therefore allow ALMA2040 to test whether GMC lifetimes are controlled primarily by internal evolution or by galactic dynamics.

This provides the direct link to stellar clumps. On galactic scales, gas surface density, rotation, and velocity dispersion constrain disk (in-)stability through the Toomre parameter \citep{Toomre1964}; on cloud scales, resolved sizes, masses, and linewidths constrain the internal dynamical state through $\alpha_{\rm vir}$ and the crossing and free-fall times. Comparing these quantities with JWST stellar morphologies and ages will test the full sequence from disk instability, through GMC formation and evolution, to the emergence and disruption of stellar clumps \citep[e.g.,][]{Linden2024}. In particular, systematic offsets between molecular gas and young stars, or trends in cloud properties with stellar evolutionary stage, will distinguish short-lived GMCs from long-lived reservoirs that survive for several internal dynamical times.
Furthermore, massive stellar clumps could also be linked to bulge formation through their migration to the galactic centers \citep{Dekel2022,Ceverino2010,Tamburello2015,ForsterSchreiber2011} and may even be responsible for the emergence of quiescent galaxies through disk stabilization \citep{Martig2009,Bournaud2016}. Measuring the life cycle of GMCs and clumps, specifically their life times, may inform about the importance of this mechanism to galaxy structure.

\section{ALMA2040: Technical Requirements}
\label{sec:technical_req}

Pursuing the above science cases requires an increased sensitivity, higher spatial resolution, and wider band coverage at sub-mm frequencies.
This translates into a next-generation sub-mm interferometer with $\approx 10\times$ increase spectral-line sensitivity to current ALMA (and $\approx 8\times$ after WSU), which can be achieved through increased collecting area and advances in receiver technology, including reduced processing overheads for large antenna numbers.

ALMA2040 will probe high-$z$ GMCs into the parameter space occupied by $z\approx0$ clouds, reaching molecular masses of $\sim10^6\,M_\odot$ and radii of $\sim40$ pc. Using Larson relations \citep{Heyer2009,Larson1981}, such clouds are expected to have $\sigma_{\rm V}\sim3.5\,\mathrm{km\,s^{-1}}$ (assuming local relation), motivating a spectral resolution of $\Delta v\lesssim8\,\mathrm{km\,s^{-1}}$ (FWHM). A spatial resolution of $\sim0.05\arcsec$ is likewise required to match the physical scales probed by JWST/NIRCam and, with lensing, to reach tens of pc. Assuming a \Cii-to-$\mathrm{M_{H_2}}$ conversion factor $\alpha_{\rm [CII]}\approx30\,\mathrm{M_\odot/L_\odot}$ \citep{Zanella2018, Madden2020, Dessauges-Zavadsky2020} and $\mu\approx20$, the expected \Cii~flux is $0.5-2\,\mathrm{mJy\,km\,s^{-1}}$ at $z\approx3-8$. For $\sigma_{\rm V}\sim3.5\,\mathrm{km\,s^{-1}}$ and a spatially and spectrally integrated $5\sigma$ detection, this corresponds to an r.m.s. of $\approx0.05-0.015\,\mathrm{mJy}$ across $z\sim3-8$.

These requirements will also enlarge the sample of massive GMCs ($\mathrm{M_{mol}\approx10^7-10^8\,M_\odot}$), currently found mainly in strongly lensed sources ($\mu\approx20-400$), by targeting the much more common $\mu<5$ population. In extreme lenses ($\mu\approx200$), ALMA2040 could reach $\mathrm{M_{H_2}\approx10^5\,M_\odot}$ at $z>1$. Whereas the required integrations with current ALMA exceed several days, a $>5\times$ and ideally $\approx10\times$ sensitivity improvement would enable detection of GMCs down to $\mathrm{M_{mol}\approx10^6\,M_\odot}$ in $\approx20$ h, and $\approx10^5\,M_\odot$ in extreme lenses with the same exposure time. This parameter space is inaccessible with current ALMA, even after WSU.

\subsection{Simulations with GHOST}
\label{subsec:simulation}

Using the {\em Generalised Hybrid Observation Support Tool} (GHOST)\footnote{\url{https://alma-ghost.jb.man.ac.uk}}, we demonstrate how an ALMA2040 concept can be used to investigate GMCs and star-forming clumps in galaxies at high redshift.

To this end, we create two complementary galaxy models.
First, we construct a model based on an observed clumpy galaxy at $z\approx1$ with a disk and $\approx50\,{\rm pc}$ GMC-like clumps \citep{Faisst2025}, then ``redshift'' it to $z=5$ using the corresponding cosmological angular-size and physical-size evolution \citep[assuming $R_e \propto (1+z)^{-1}$, e.g.,][]{Ormerod2024}.
Second, we construct a model with \texttt{astropy.models} resembling galaxies observed in large ALMA programs such as ALPINE \citep{Lefevre2019,Bethermin2020,Faisst2020}, CRISTAL \citep{HerreraCamus2025}, and REBELS \citep{Bouwens2022}; it contains a bulge and $15$ clumps with different brightnesses. In both cases, the total flux is normalized to $1\,{\rm mJy}$ (corresponding to a total infrared luminosity of $\approx3\times10^{11}\,{\rm L_\odot}$) and the intrinsic pixel scale is $4\,{\rm mas}$.

We then run the models through GHOST using the \Cii~line at rest-frame $345\,{\rm GHz}$ (band 7) at $z=5$, assuming $1\,{\rm mm}$ PWV and a source declination of $2\,\mathrm{deg}$:
\begin{itemize}
    \item ALMA 12m C-4 configuration, Briggs ($r=0.5$) weighting ($\theta\approx 0.33\arcsec$) at 5 hours time on source;\vspace{-3mm}
    \item ALMA 12m C-8 configuration, Briggs ($r=0.5$) weighting ($\theta\approx 0.04\arcsec$) at 12 hours time on source;\vspace{-3mm}
    \item ALMA2040 Intermediate configuration, Briggs ($r=0.5$) weighting ($\theta\approx 0.05\arcsec$) at 5 hours time on source;
\end{itemize}

The results based on the empirical galaxy model are shown in Fig.~\ref{fig:ghost}A, while the results from the simulated model galaxy are shown in Fig.~\ref{fig:ghost}B.

We also simulate the corresponding ELT/MICADO $K_s$-band images, considering the Multi-Conjugate Adaptive Optics mode of MORFEO, the adaptive optics (AO) module that will feed MICADO. We use the ScopeSim simulator\footnote{https://scopesim.readthedocs.io/en/latest/getting\_started.html}, median atmospheric conditions, and a realistic PSF from the TipTop simulator\footnote{https://astro-tiptop-services.github.io/astro-tiptop-services/}, using a real asterism of three natural guide stars surrounding the Sunburst arc \citep{Rivera-Thorsen2017}.

Simulations show that ALMA2040 resolves ISM structures far more efficiently than current ALMA configurations, including the post-WSU update, within practical observing times. MICADO on ELT will resolve stellar clumps, while ALMA2040 will provide complementary views of cold molecular gas. Together, they will enable spatially resolved comparisons of GMCs, stellar clumps, and their evolutionary state (see also Sect.~\ref{subsec:synergies}).

\section{Synergies with current and future facilities}
\label{subsec:synergies}

By the 2040s, observations with JWST and large surveys with {\em Euclid} or the {\em Nancy Grace Roman} space telescope will have provided large, statistically representative samples of clumpy galaxies at $z>3$.
Similarly, the ELT will deliver unprecedented spatial resolution for measuring cluster sizes and performing optical to near-infrared spectroscopy in galaxies with mild magnification and down to $\approx20$ pc scales in systems with $\mu\approx5$. This will cover UV and optical emission lines to measure the ionized ISM including outflows, metallicity, and stellar populations properties.
SKAO will add information on atomic gas from star-forming regions at high redshift at comparable scales.
ALMA2040 will fit perfectly in this landscape; it will provide the complementary census of molecular gas and dust, making it possible to connect the stellar, atomic, and molecular phases across the scales from GMCs to galaxies and across the evolution from Cosmic Noon to the Epoch of Reionization.

\section*{Acknowledgements}
This work was developed as part of the broader ALMA2040 community effort. We acknowledge the Lorentz Center and NOVA for supporting the community workshops and coordination activities that helped initiate and shape the ALMA2040 White Paper process.\\

\bibliographystyle{apj_mod}
{\bibliography{bibliography.bib}}

\newpage
\section*{Author Affiliations}
{
\footnotesize{
$\dagger$These authors contributed equally to this work\vspace{0.5em}\\
\noindent $^1$ Istituto Nazionale di Astrofisica (INAF), Via Gobetti 93/3, 40129 Bologna, Italy, \\
$^2$ IPAC, California Institute of Technology, 1200 E. California Blvd. Pasadena, CA 91125, USA, \\
$^3$ Millennium Nucleus for Galaxies (MINGAL), Av. Ej\'ercito 441, Santiago 8370191, Chile \\
$^4$ Departamento de Astronom\'ia, Universidad de Concepci\'on, Barrio Universitario, Concepci\'on, Chile \\
$^5$ Princeton University, Princeton, NJ 08544, USA\\
$^6$ SKA Observatory, Jodrell Bank, SK11 9FT, UK\\
$^7$ Instituto de Estudios Astrof\'{\i}cos, Facultad de Ingenier\'{\i}a y Ciencias, Universidad Diego Portales, Av. Ej\'ercito 441, Santiago, Chile\\
$^8$ DARK, Niels Bohr Institute, University of Copenhagen, Jagtvej 155A 2200 Copenhagen, Denmark\\
$^{9}$ Department of Physics and Astronomy, Chalmers University of Technology, SE-412 96 Gothenburg, Sweden\\
$^{10}$ Max-Planck-Institut f\"ur Astrophysik, Karl-Schwarzschild-Str. 1, D-85748 Garching, Germany\\
$^{11}$ INAF -- IASF Milano, via A. Corti 12, I-20133 Milano, Italy\\
$^{12}$ Leiden Observatory, Leiden University, PO Box 9513, NL-2300 RA Leiden, The Netherlands\\
$^{13}$ Faculty of Mathematics and Physics, Jadranska ul. 19, SI-1000 Ljubljana, Slovenia\\
$^{14}$ Instituto de F\'{i}sica y Astronom\'{i}a, Universidad de Valpara\'{i}so, Gran Breta\~{n}a 1111, Valpara\'{i}so 2360102, Chile\\
$^{15}$ Sorbonne Université, UPMC Paris 6 and CNRS, UMR 7095, Institut d’Astrophysique de Paris, 98b bd. Arago, 75014 Paris, France\\
$^{16}$ D\'epartement d’Astronomie, Universit\'e de Gen\`eve, Chemin Pegasi 51, 1290 Versoix, Switzerland\\
$^{17}$ Institute of Astrophysics, Foundation for Research and Technology-Hellas (FORTH), Heraklion, 70013, Greece\\
$^{18}$ School of Sciences, European University Cyprus, Diogenes Street, Engomi, 1516 Nicosia, Cyprus\\
$^{19}$ ENS de Lyon, Universit\'e Lyon 1, CNRS, CRAL, UMR 5574, 69007 Lyon, France\\
$^{20}$ Universit\'e de Strasbourg, CNRS, Observatoire astronomique de Strasbourg, UMR 7550, F-67000 Strasbourg, France
$^{21}$ SISSA, Via Bonomea 265, 34136 Trieste, Italy\\
$^{22}$ INAF-Italian National Institute of Astrophysics, Observatory of Trieste, Via G. Tiepolo 11, 34143 Trieste, Italy\\
$^{23}$ Institute for Fundamental Physics of the Universe, Via Beirut 2, 34151 Trieste, Italy\\
$^{24}$ Department of Astronomy, University of Michigan, 1085 South University Avenue, Ann Arbor, MI 48109, USA\\
$^{25}$ Instituto de F\'isica Fundamental, CSIC, Calle Serrano 123, 28006 Madrid, Spain\\
$^{26}$ Kapteyn Astronomical Institute, University of Groningen, Landleven 12, 9747 AD, Groningen, The Netherlands\\
$^{27}$ Department of Physics and Astronomy, Texas A\&M University, 4242 TAMU, College Station, TX 77843-4242, US \\
$^{28}$ European Southern Observatory (ESO), Karl-Schwarzschild Str. 2, 85748 Garching, Germany
}
}
\end{document}